\documentclass[12pt,a4paper,final]{iopart}

\pdfoutput=1 

\usepackage{iopams}
\usepackage{graphicx}
\usepackage[breaklinks=true,colorlinks=true,linkcolor=blue,urlcolor=blue,citecolor=blue]{hyperref}

\usepackage[T1]{fontenc} 
\usepackage[utf8]{inputenc}

\usepackage{color}

\newcommand{\jat}[1]{{\color[rgb]{0,0.0,0}#1}}

\newcommand{\ddt}[1]{{\color[rgb]{0,0.0,0}#1}}

\newcommand{\rdt}[1]{{\color[rgb]{0.0,0,0.0}#1}}

\newcommand{\cct}[1]{{\color[rgb]{0.0,0,0.0}#1}}
\newcommand{\mkt}[1]{{\color{black}#1}}

\newcommand{\de}{\delta}

\newcommand{\beq}{\begin{equation}}
\newcommand{\eeq}{\end{equation}}

\newcommand{\dd}{\mathrm{d}}
\newcommand{\cH}{\mathcal{H}}

\newcommand{\gev}{{\em gevolution}}

\renewcommand{\dot}[1]{{#1}'}

\begin{document}

\title[Backreaction in relativistic cosmological simulations]{Safely smoothing spacetime:\\ Backreaction in relativistic cosmological simulations}

\author{\jat{Julian Adamek}$^{1,2}$, \cct{Chris Clarkson}$^{2,3,4}$, \ddt{David Daverio}$^5$, \rdt{Ruth~Durrer}$^6$, \mkt{Martin Kunz}$^6$}
\address{$^1$ Laboratoire Univers et Th\'eories -- UMR 8102, Observatoire de Paris, CNRS, PSL Research University, 5 Place Jules Janssen, 92195 Meudon CEDEX, France}
\address{$^2$ School of Physics \& Astronomy, Queen Mary University of London, 327 Mile End Road, London~E1~4NS, United Kingdom}
\address{$^3$ Department of Mathematics \& Applied Mathematics, University of Cape Town, South Africa}
\address{$^4$ Department of Physics and Astronomy, University of the Western Cape,  South Africa}
\address{$^5$ Centre  for  Theoretical  Cosmology,  Department  of  Applied  Mathematics  and  Theoretical  Physics,
University  of  Cambridge,  Wilberforce  Road,  Cambridge~CB3~0WA,  United  Kingdom}
\address{$^6$ D\'epartement de Physique Th\'eorique \& Center for Astroparticle Physics, Universit\'e de Gen\`eve, Quai E.\ Ansermet 24, CH-1211 Gen\`eve 4, Switzerland}
\ead{julian.adamek@qmul.ac.uk, chris.clarkson@qmul.ac.uk, dd415@cam.ac.uk, ruth.durrer@unige.ch, martin.kunz@unige.ch}

\date{\today}

\begin{abstract}

A persistent theme in the study of dark energy is the question of whether it really exists or not. It is often claimed that we are mis-calculating the cosmological
model by neglecting the effects associated with averaging over large-scale structures. In the Newtonian approximation this is clear: there is no effect. Within the
full relativistic picture this remains an important open question, owing to the complex mathematics involved. We study this issue using numerical N-body simulations 
which account for all relevant relativistic effects without any problems from shell crossing. In this context we show for the first time that the backreaction
from structure can differ by many orders of magnitude depending upon the slicing of spacetime one chooses to average over. In the worst case, where smoothing is carried
out in synchronous spatial surfaces, the corrections can reach ten percent and more. However, when smoothing on the constant time hypersurface of the Newtonian gauge,
backreaction contributions remain 3-5 orders of magnitude smaller. 

\end{abstract}



\maketitle

\section{Introduction}
\label{sec:intro}

The expansion of the Universe is accelerating. This surprising finding has led to the Nobel Prize in 2011~\cite{Riess:1998cb,Perlmutter:1998np,Schmidt:1998ys} and has been confirmed with many other data since, e.g.\ \cite{Bassett:2003vu,Eisenstein:2005su,Sherwin:2011gv,Ade:2015xua}.
Within general relativistic Friedmann-Lema\^\i tre-Robertson-Walker (FLRW) cosmology, the simplest solution points to a cosmological constant, which now underpins the $\Lambda$-cold-dark-matter (LCDM) model. 
 Even though a cosmological constant provides an excellent fit to the data (see e.g.,~\cite{Ade:2015xua,Aghanim:2015xee,Ade:2015rim}), it requires an extremely fine tuned non-vanishing vacuum energy, which remains unexplained. Other models include modifications to general relativity on large scales, or dynamical scalar fields, see~\cite{Copeland:2006wr,Durrer:2007re,Clifton:2011jh,Amendola:2016saw} for a review. All models tend to require fine tuning to explain why acceleration is beginning now, when we happen to observe it, this is called the coincidence problem.
 
A different idea arises from the question whether the structures present in the Universe might affect measurements in such a way that we infer accelerated expansion once we interpret these measurements within the framework of an FLRW model. A particularly attractive feature of this concept is that it would solve the coincidence problem, explaining why acceleration begins roughly at the same time that non-linear structures form.
Furthermore, even if backreaction does not produce a fully-fledged dark energy model, perhaps it can alter quantities such as the spatial curvature by a significant  amount. It is a matter of some importance to quantify these effects in more detail {as they may bias parameter inferences from observations.}

Whether the backreaction idea works
remains a subject of considerable debate~-- see~\cite{Buchert:2007ik,Rasanen:2011ki,Buchert:2011sx,Clarkson:2011zq,Clifton:2013vxa} for reviews and~\cite{Bolejko:2016qku} for a survey of cosmologists' opinions. Within the standard Newtonian approximation, the answer is `no', unless peculiar boundary conditions are imposed~\cite{Buchert:1999er,Buchert:2001sa} (see \cite{Kaiser:2017hqn,Buchert:2017obp} for a recent revival). In this paper we do not study the related problem of fitting observables with an inhomogeneous background, or with average light propagation in an inhomogeneous model, but we will comment on it in our conclusions.

Within the framework of general relativity \mkt{(GR)} the question of backreaction is considerably more subtle than in Newtonian gravity, and the debate, in some circles, is intense~\cite{Buchert:2015iva,Green:2015bma}. In essence the problem arises because any average relies on knowing the fully non-linear spacetime geometry so we cannot a priori define quantities like a mean
energy density; the Einstein equations are non-linear; and,  averages of tensors are not well defined. Many ideas have been proposed, with conclusions drawn
depending on the approach taken~\cite{Clarkson:2011zq}. Analyses based on standard perturbation theory typically give a small~-- sub-percent~-- correction to the background~\cite{Rasanen:2004sa,Clarkson:2009hr}. It does depend precisely on the quantity being averaged~-- certain quantities are divergent, see for
example~\cite{Clarkson:2011uk}, while many average quantities depend on the choice of hypersurface~\cite{Ishibashi:2005sj,Clarkson:2009hr,Gasperini:2009mu,Marozzi:2010qz,Umeh:2010pr}. Another significant complication~-- which we address here~-- is 
that perturbative approaches rely on a fluid approximation which breaks down when shell crossing happens, rendering conclusions tentative. Alternative approaches typically use highly simplified exact solutions to determine the feasibility of 
backreaction as an effective dark energy~\cite{DiDio:2011gf,Clifton:2012qh,Bentivegna:2012ei,Yoo:2013yea}. Both approaches have their drawbacks (quasi-Newtonian versus over-simplified) and it is difficult to draw definitive conclusions. 

A new approach is now available with the advent of cosmological N-body simulations which incorporate all relevant general relativistic effects~\cite{Adamek:2014xba,Adamek:2015eda,Adamek:2016zes}. This allows us to approach these questions from another point of view. We argue that the question is not whether there are time-slicings (observers) for which there is significant backreaction but whether there are
slicings such that backreaction is small or even vanishes. In particular, one important criticism
of some approaches that argue for a large backreaction effect is that the averages are performed on hypersurfaces orthogonal to observers comoving with the matter
flow~\cite{Ishibashi:2005sj,Umeh:2010pr}, which also is a natural slicing for many recent studies based on numerical relativity, e.g. \cite{Bentivegna:2015flc,Mertens:2015ttp,Giblin:2015vwq,Giblin:2017juu}. But other slicings, such as the harmonic slicing, have also been used, see e.g.\ \cite{Macpherson:2016ict}. In fact, after our work was completed, \cite{Macpherson:2018akp} conducted a similar study using a slicing that is quite close to the one of Poisson gauge, finding good agreement with our results. Apart from the slicing issue the main difference between the numerical approaches concerns the matter model. In the numerical relativity community the matter is modelled as a perfect fluid which limits these studies to either very large scales or special situations. Our relativistic simulations employ the N-body method that \mkt{can handle} the problem of shell-crossing and gives us access to a much larger dynamical range in the non-linear regime. While being fully relativistic, our framework breaks down in the regime of strong gravity, e.g.\ close to black holes. This regime is never probed in cosmological simulations. \mkt{Thanks to the focus on the cosmologically relevant situation of weak gravitational fields we are also able to use simulation grids of size $2048^3$ and more, much larger that what could so far be achieved with other cosmological full-GR simulations.}

For the purpose this work we say that a universe is called a {FLRW} universe with small perturbations if there {\it exist} coordinates such that the metric fluctuations, averaged over 
sufficiently large scales of order 1\,Mpc are small.  Intuitively, the reason that we can average over such a scale even if inside there may be black holes is
the following: At a distance $r\simeq$ 1\,Mpc, the potential of a $10^{13}M_\odot$ object is of the order of $R_S/r\simeq 10^{-6}\ll 1$ (this is the monopole part of the potential, and we know that
the potential of the higher multipoles generically decays even faster, like $r^{-(1+\ell)}$ and hence is even less relevant). At the scale of 1\,Mpc it should therefore
no longer make a difference whether the field was generated by a compact, strong-field source or by a somewhat more extended source that generates only weak gravitational
fields everywhere inside the domain. In addition, we note that the strong equivalence principle guarantees that a galaxy that contains some black holes falls in the same way as a galaxy that contains only stars.

In this paper we show that backreaction is small in the sense that there exists a time slicing such that it is small. We identify this as the time slicing of Poisson gauge (also called 
Newtonian or longitudinal gauge in the context of scalar perturbations). We also show that in another slicing which is often used, namely geodesic slicing which is related to comoving gauge, backreaction can become large.

In the remainder of this paper we first discuss backreaction within a perturbative approach {in Section 2}. Even though this is of course not sufficient, it gives us important indications
on the slicings within which backreaction becomes strong. {In Section 3 we} then show numerical results from the fully relativistic weak field N-body code~\gev~\cite{Adamek:2015eda,Adamek:2016zes}. Finally we discuss our results and conclude.
\vspace{0.2cm}

\section{Quantifying backreaction}

{Even though the goal of this paper is a numerical study of backreaction, let us first discuss backreaction with a perturbative approach to gain some analytical insight.}
Backreaction is typically quantified by averaging various scalars and comparing them {to their analogs in a perfectly smooth spacetime} -- usually an FLRW model with similar matter content. In the case of perturbation theory but also for numerical simulations that start from appropriate initial data there is a well defined background to compare to. Here we  consider the expansion rate for different families of observers. We consider observers, with 4-velocity $u^\mu$, in comoving gauge in which the equal-time hypersurfaces coincide with the matter rest frame. We also consider another frame, $n^\mu$, which is the {normal to the equal time hypersurfaces in} Poisson gauge. 

Including only scalar, linear perturbations the metric in a generic gauge is \cite{Kodama:1985bj}
\begin{eqnarray}
 \dd s^2 &=& g_{\mu\nu} \dd x^\mu \dd x^\nu = a^2(\tau) \Bigl[-\left(1+2A\right) \dd\tau^2 - 2 B_{,i} \dd x^i \dd\tau \Bigr.\nonumber\\ \Bigl.&&+ \left(1+2H_L\right) \delta_{ij} \dd x^i \dd x^j + 2({H_T}_{,ij} - \frac{1}{3} \delta_{ij} \Delta H_T)\dd x^i \dd x^j\Bigr] \, ,
\label{eq:metric-gen}
\end{eqnarray}
where $a$ is a background scale factor, $\tau$ is conformal time, and $x^i$ are comoving Cartesian coordinates.
To first order in perturbation theory 
the expansion rate normal to the $\{\tau={\rm const.}\}$ hypersurfaces is~\cite{durrer2008cosmic} 
\beq\label{e:theta-gen}
\theta =3H\left(1-A+\cH^{-1}\dot H_L -\frac{1}{3}\cH^{-1} \Delta B\right)\,,
\eeq
which of course only depends on the temporal gauge choice and not on the spatial one.
Here $H$ and  $\cH$ are the physical and conformal Hubble parameter, $H=\dot a/a^2$ and  $\cH=\dot a/a$. A prime denotes a derivative with respect to conformal time $\tau$.
From this expression one immediately infers that choosing a temporal gauge (or more precisely, the corresponding time slicing) where $A-\cH^{-1}\dot H_L +\frac{1}{3}\cH^{-1} \Delta B=0$ there is no backreaction at first order in perturbation theory.

We now consider the time slicing given by Poisson gauge, in which $B=H_T=0$, and $A=\Psi$ and $-H_L=\Phi$ are the usual Bardeen potentials. The expansion rate in this case reduces to
\beq\label{e:theta-P}
\theta^{(P)} =3H\left(1-\Psi-\cH^{-1} \dot\Phi\right)\,.
\eeq
Introducing $\chi=  \Phi - \Psi$ we find that a correction to the background expansion rate in any sub-box of our simulation is given by (see also section 5.3 of \cite{Adamek:2016zes})
\beq
\tilde{\cH} = \cH \left( 1 - 2 \bar{\Phi} + \bar{\chi} - a \frac{d\bar{\Phi}}{da} \right)\, .
\eeq
Here the over-bar denotes an average over a sub-volume and we fix conformal time $\tau$ for all boxes by the condition that for the full box $\bar\chi=0$, but not on sub-volumes. This fixes the residual gauge-freedom of Poisson gauge~\cite{Adamek:2016zes}. The left-hand side is evaluated at a perturbed scale factor $\tilde{a} = a (1-\bar\Phi)$, so that we finally find a perturbation of $\cH$, evaluated at the perturbed redshift $\tilde{z}=\tilde a^{-1}-1$, of
\beq
\left(\frac{\Delta\cH}{\cH}\right)^{(P)} = - 2 \bar{\Phi} + \bar{\chi} - \tilde a \frac{d\bar{\Phi}}{d\tilde a} + \frac{\tilde{a}}{\tilde{\cH}} \frac{d\tilde{\cH}}{d\tilde{a}} \bar{\Phi} \, . 
\label{eq:tn1}
\eeq
{This equation is correct in linear perturbation theory. However, when using N-body simulations to compute $\Phi$ and $\chi$, as will be done in the next section, the difference to the fully non-perturbative result is of quadratic order in $\Phi$ and $\chi$ and hence the relative change is not larger than $10^{-4}$.} {Therefore (\ref{eq:tn1}) is a good approximation to the full GR result in Poisson gauge. }

Another possible coordinate choice is comoving gauge where the equal time hypersurfaces coincide with the matter rest frame. This is well defined as long as we can neglect the fluid vorticity. 
For cold dark matter (CDM) perturbations the comoving gauge is also synchronous so that $A=0$ in this gauge. In linear perturbation theory we can use the conservation equation to rewrite (\ref{e:theta-gen}) as
\beq
\theta^{(u)} = \frac{3 \cH}{a} \left(1- a \frac{d \delta}{da} \right)
\label{eq:tu2}
\eeq
for the longitudinal gauge density perturbation $\delta$. We can define a corresponding perturbation of the conformal Hubble parameter in CDM comoving gauge,
\beq
\left(\frac{\Delta\cH}{\cH}\right)^{(u)} = \bar{H}_L - a \frac{d\bar{\delta}}{da} - \frac{\tilde{a}}{\tilde{\cH}} \frac{d\tilde{\cH}}{d\tilde{a}} \bar{H}_L \, . 
\label{eq:tu3}
\eeq
{Again, this expression is correct in linear perturbation theory. Contrary to Eq.\ (\ref{eq:tn1}), higher-order corrections to this equation could be of order $\delta^2$, which is large once structure formation becomes non-linear.
Therefore Eq.\ (\ref{eq:tu3}) is a bad approximation to $\Delta\cH/\cH$ in comoving gauge, while (\ref{eq:tn1}) should be very good.} 

Comparing expressions (\ref{eq:tn1}) and (\ref{eq:tu3}) we further notice that the first one remains always small, of the order of the gravitational potentials. The second one however is of the order of $ ad\delta/da \approx \delta$ which becomes of order unity and more at late times. This quantity cannot be `compensated' by the metric potentials in comoving gauge as within linear perturbation theory and also in the Newtonian limit metric potentials are of the order $(\cH/k)^2\delta$, hence much smaller than density perturbations on sub-horizon scales. In a cosmological setting, this relation remains also valid in full GR on sub-horizon scales, so that the fully non-linear metric potentials that define the Poisson gauge metric remain small at all times. This makes the Poisson gauge so well suited for cosmological N-body simulations.\footnote{It is of course possible to define coordinate systems such that the metric deviates strongly from the Friedmann one. But relativistic numerical simulations~\cite{Adamek:2015eda,Adamek:2016zes} have shown that the converse is also true. We can define coordinates in e.g. Poisson gauge, for which metric perturbations remain small even when density fluctuations become large.} 

The volumes of the the first case are the equal-time hypersurfaces of Poisson gauge, describing a Newtonian frame, and the expansion rate $\theta^{(P)}$ describes their deformation. The volumes in the second case are ``attached'' to the particles in the simulation and follow their expansion and collapse\footnote{In general this is
no longer a well-defined prescription as soon as there is shell crossing. At this point also vorticity develops and comoving gauge breaks down. In our simulations this
breakdown is avoided since we employ a coarse-graining to relatively large scales defined through our sub-volume averages.}. In the first case there is in general a particle flux across the boundaries of sub-volumes, in the second case there is no such flux by definition. It is sometimes argued that the latter is more relevant for observations since observers are typically embedded in matter. However, we would rather argue the opposite by pointing out that observations are typically taken along null directions, and the relation between volumes and their \textit{appearance} on the past light cone is much simpler in the former case -- an observer indeed sees matter falling towards each other.
\vspace{0.2cm}

\section{Numerical analysis}

In this section we go beyond perturbation theory and investigate $\Delta\cH/\cH$ with numerical simulations. For this purpose we use the relativistic N-body code \gev\ \cite{Adamek:2015eda,Adamek:2016zes}. Even though also \gev\ neglects certain terms that become important when gravity becomes strong, for example near black holes, for cosmological applications and in the gauge used by \gev\, these higher-order terms of the weak-field expansion remain a small and numerically irrelevant relative correction of the order of $10^{-4}$.

In practical terms, we extract the quantities given in Eq.\ (\ref{eq:tn1}) and the dominant term of (\ref{eq:tu3}) from numerical simulations and compute $\Delta\cH/\cH$. While, as discussed in the previous section, expression (\ref{eq:tn1}) is an excellent approximation to the full non-perturbative $\Delta\cH/\cH$ in the time slicing of Poisson gauge, a corresponding statement is probably not true for (\ref{eq:tu3}).

We consider a full simulation volume, and divide it into sub-boxes to quantify backreaction in each.
We performed several simulations with comoving linear box sizes of $2048$ Mpc/$h$ and $512$ Mpc/$h$ and a grid size of $2048^3$. All simulations {use} $\Omega_m h^2 = 0.142412$, $A_s = 2.215 \times 10^{-9}$ and $n_s = 0.9619$. For the LCDM simulations we set $h=0.67556$ which implies $\Omega_\Lambda = 0.6879$, while for the Einstein -- de Sitter (EdS) simulations we use $h= 0.3774306$ so that $\Omega_m = 1$. The initial spectra are generated with CLASS \cite{Blas:2011rf} at a redshift of $z=100$, which is also the starting redshift for the simulations.

\begin{figure}[t]
\centerline{\includegraphics[width=0.66\textwidth]{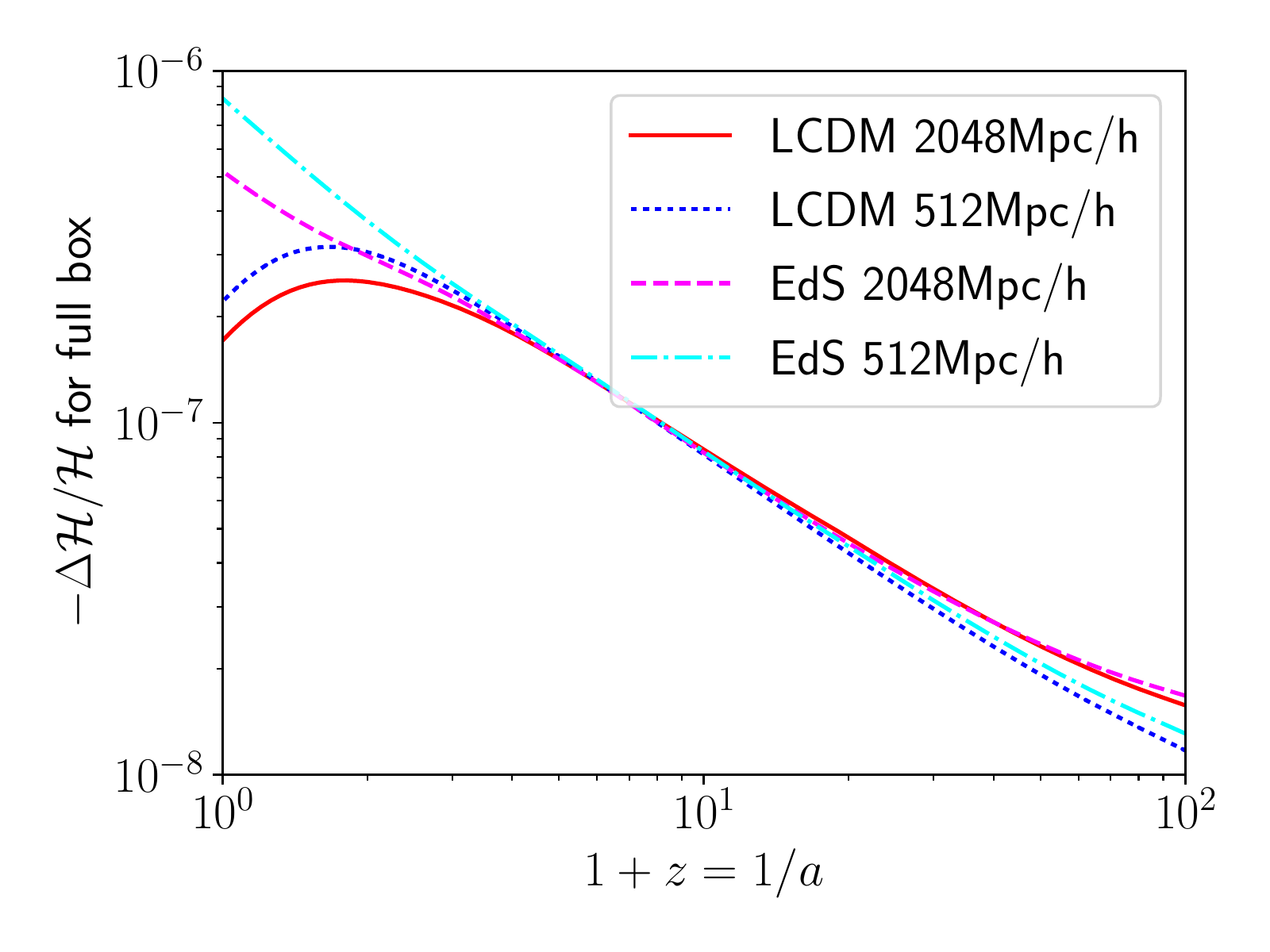}}
\caption{\label{fig:fullboxbr} We show $\Delta\cH/\cH$ in the time slicing of Poisson gauge for two different box sizes in an EdS universe and in a LCDM universe, averaging over the full simulation box. Backreaction slows down the expansion, and even though the absolute value is growing, it remains smaller than $10^{-6}$ at all times.}
\end{figure}

\begin{figure}[t]
\centerline{\includegraphics[width=0.66\textwidth]{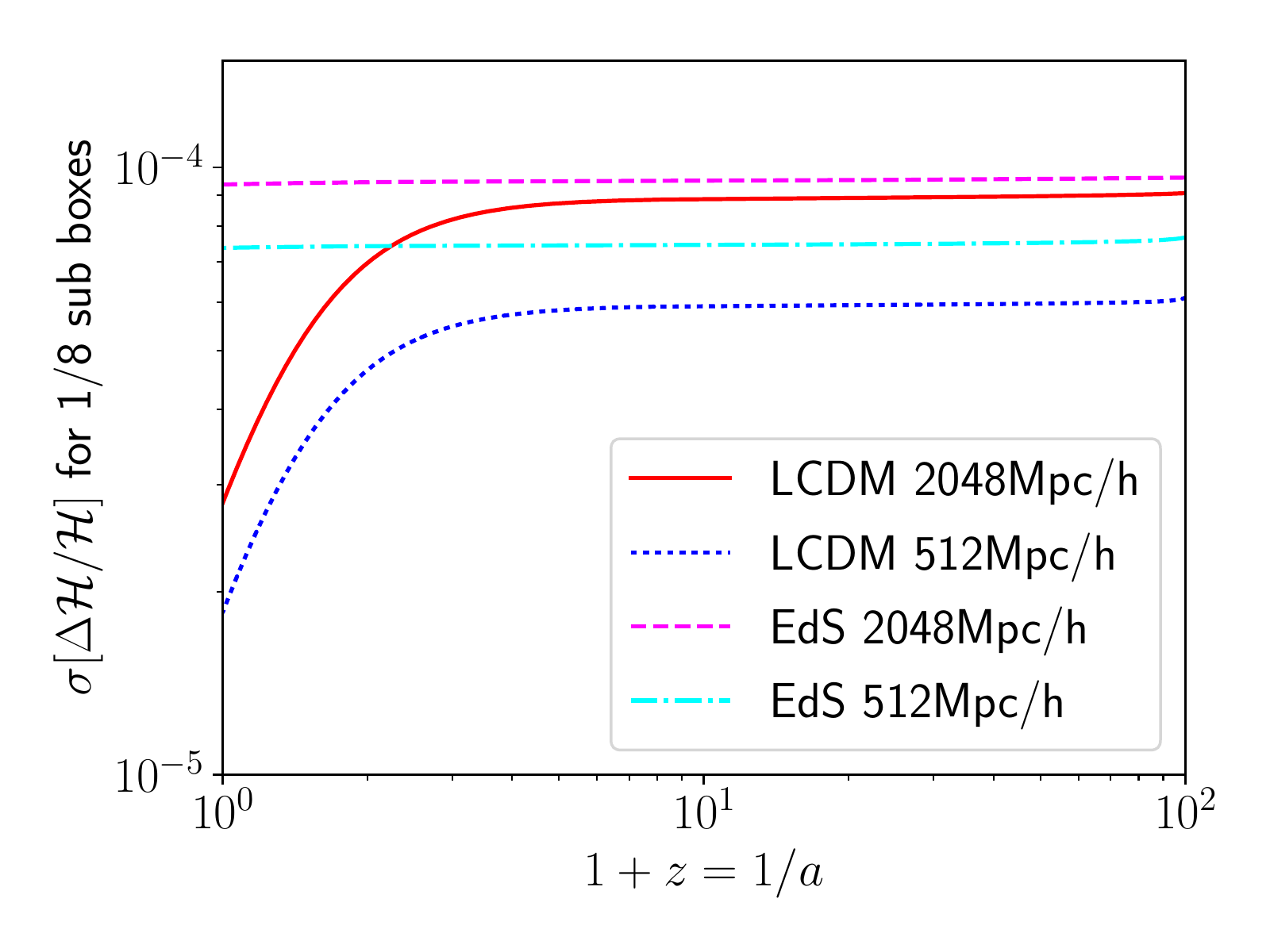}}
\caption{\label{fig:subboxbr} We show the standard deviation of $\Delta\cH/\cH$ in the time slicing of Poisson gauge for two different box sizes in an EdS universe and in a LCDM universe, averaging over sub boxes of linear size of 1/8 of the full box. The mean value over all boxes is by construction equal to the average over the full box shown in Fig.\ \ref{fig:fullboxbr}. The standard deviation is more than two orders of magnitude larger, but still less than $10^{-4}$. It remains relatively constant over time in the EdS universe and decays during $\Lambda$-domination in a LCDM universe.}
\end{figure}

\begin{figure}[t]
\centerline{\includegraphics[width=0.66\textwidth]{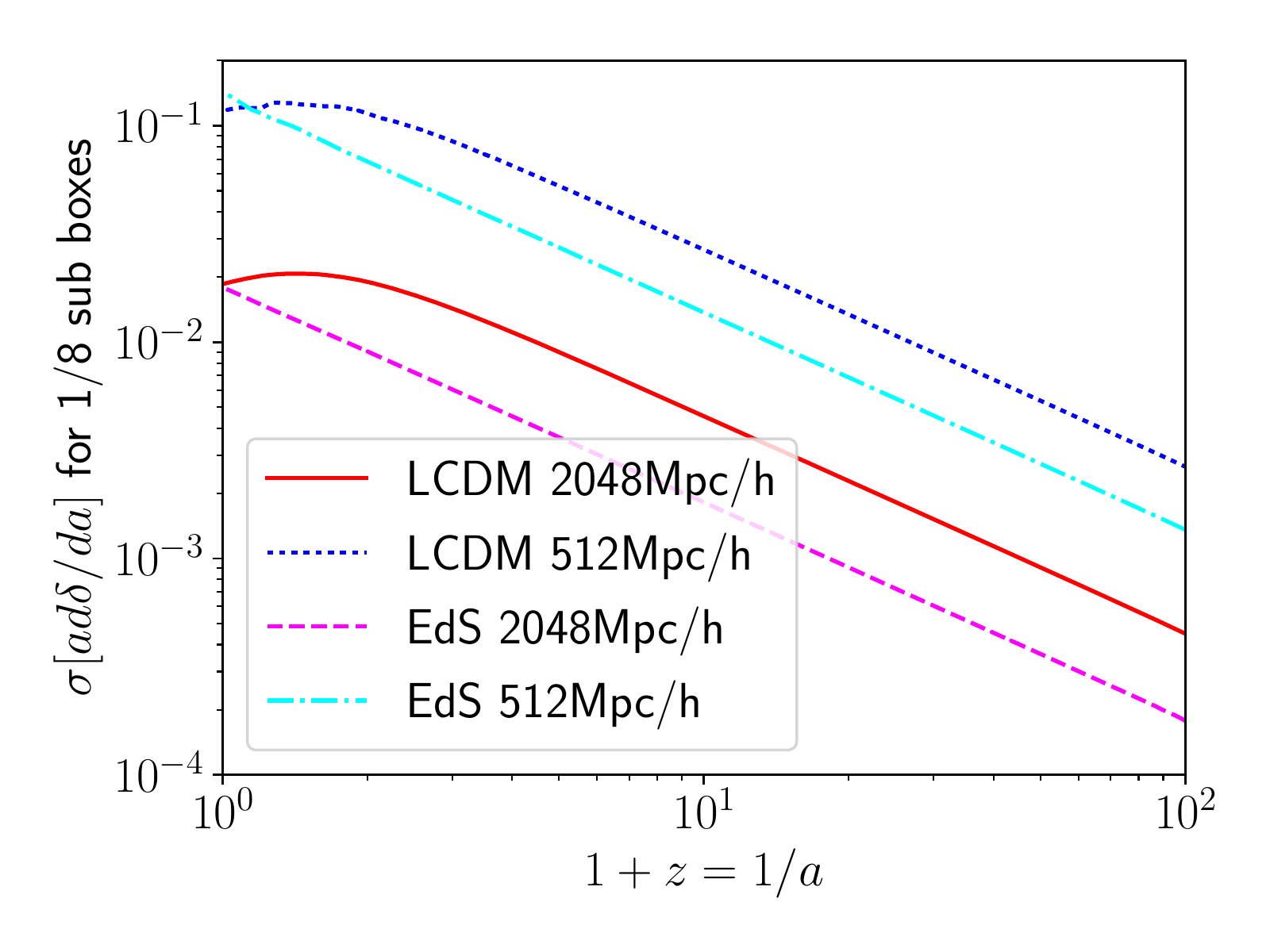}}
\vspace{-10pt}\caption{\label{fig:dop} We show the standard deviation of the dominant contribution, $a d\delta/da$, to {the linearised expression for} $(\Delta\cH/\cH)^{(u)}$ for two different global box sizes in sub boxes of 1/8th linear size, in an EdS universe and in a LCDM universe. For the small sub-boxes of size 64 Mpc/$h$ the backreaction \cct{can} reach nearly 15\%.}
\end{figure}

In the Poisson gauge slicing, backreaction remains below $10^{-4}$ even in relatively small sub-boxes of size 64 to 256 Mpc/$h$, see Fig.\ \ref{fig:subboxbr}. It
depends only weakly on the size of the box and is redshift independent for an EdS universe. This agrees with the linear perturbation theory expression (\ref{eq:tn1})
as the gravitational potentials remain constant. In a LCDM universe the potentials decay at late time, and so does  backreaction.

It is also intriguing that the backreaction in the smaller sub-box is somewhat smaller. This might be linked to the slightly red primordial power spectrum of  perturbations and to the fact that on small scales, below about 1 Gpc/$h$, the gravitational potential decreases rapidly so that the smaller box sees less overall power. However, $\Delta\cH/\cH$ is  also affected by a significant realization noise of up to a factor two so that in any case the curves in Figs.\ \ref{fig:fullboxbr} and \ref{fig:subboxbr} should be taken as indicating the order of magnitude of the effect.

Interestingly, even though the backreaction in the full box, shown in Fig.\ \ref{fig:fullboxbr}, is much smaller than the one in sub-boxes, it is time dependent and growing with time. It is this backreaction that was studied in \cite{Adamek:2014gva} in a plane symmetric relativistic and in a post-Newtonian context. With our new 3D relativistic \mkt{$N$-body} setup we find a good qualitative agreement with those previous results. \mkt{For earlier perturbative calculations see e.g.\ \cite{Rasanen:2003fy,Kolb:2004am}}.

In~Fig.\ \ref{fig:dop} we show the dominant contribution to Eq.\ (\ref{eq:tu3}), $a d\bar\delta/da$, which governs backreaction in a time slicing adapted to comoving gauge 
for sub-boxes of size 64  Mpc/$h$ (blue dotted and cyan dash-dotted) and 256  Mpc/$h$ (red solid and magenta dashed) -- for the full box $\delta$ vanishes at first order. There is clearly a big difference between the backreaction in the respective time slicings of Poisson gauge and comoving gauge. In the latter backreaction could become order 15\% in the sub-boxes of size 64 Mpc/$h$. The smaller the box size the larger  backreaction.  On even smaller scales, where 
$\bar\de>1$, backreaction could become of order unity in this slicing.
\vspace{0.2cm}

\section{Conclusions}

The important message of this paper is that the backreaction in the expansion rate depends on the choice of slicing. 
In order to answer the physically relevant question to which extent backreaction can bias our observations one should compute observables or at least choose a slicing that can easily be related to observations. Cleary, \rdt{given a spacetime,} observables can be computed in an arbitrary slicing, and the results will be identical and independent of the coordinate system used. \rdt{However, smoothing a spacetime transforms it into a different one which has no fluctuations on the hypersurface chosen to smooth over. The resulting spacetime depends of course on the choice of  the smoothing hypersurface. The interesting question is, how close an observable in the fluctuating universe is to the smoothed one. We have not answered this question here. However we have shown that in Poisson gauge} the deviations from linear perturbation theory in $\Delta\cH/\cH$\jat{, which is however \textit{not} an observable, remain small, and at the same time} the deformation of the light cone, \jat{which finally connects this quantity to observables, is likewise small.}
\rdt{This hints, that when smoothing over Poisson gauge hypersurfaces,  observables in the smoothed spacetime may remain close to the fluctuating ones.}

It is often argued that observers are made up of baryons and therefore comoving synchronous gauge provides the correct slicing to use (neglecting velocity bias).
However, during the non-linear evolution of gravitational clustering, particles undergo shell crossing and the comoving gauge breaks down. While we can consider the average velocity field in each simulation cell, this does not in general lead to a well-defined slicing as this coarse-grained velocity field has significant vorticity \cite{Jelic-Cizmek:2018gdp}.
For this reason, this gauge
is not well suited for quantifying the observed backreaction in the late universe when most of the matter is in the form of compact objects in highly random motion,
separated by vast stretches of empty space. Since the gravitational fields of these objects are weak, quasi-Newtonian, Poisson gauge is ideally suited for studying the non-linear
evolution in this era. As we have shown here, in the time slicing of Poisson gauge backreaction remains small, well below 1\% for the Hubble parameter, at all times, and for different
sub-box sizes, for both EdS and LCDM cosmologies. This was expected from theoretical considerations, see e.g.\ \cite{Green:2010qy,Rasanen:2011bm}, but we
quantify this statement numerically.

It is important to note that  in Newtonian gravity backreaction is a pure boundary term~\cite{Buchert:1995fz,Buchert:1999pq} and thus absent in Newtonian N-body simulations with periodic boundary conditions. Although \gev\ also uses periodic
boundary conditions, it sees a non-zero backreaction as it works with full general relativity in the weak-field regime. The periodic boundary conditions however do impose
the global constraint that the \mkt{particle number in the box is fixed. At first order this corresponds to a constant energy density and hence, for our initial conditions, to a vanishing 
average spatial curvature $\bar{R}^{(3)}=0$}. 
When we consider backreaction in the sub-boxes we no longer have this constraint, and the periodicity of the boundary conditions is also removed. To the extent that the sub-boxes evolve as separate universes, they are initialized with slightly varying cosmological parameters, including some non-vanishing average spatial curvature. 
However, the results shown in Fig.~\ref{fig:subboxbr} can not be explained in this way since the effect of spatial curvature would have a strong redshift dependence.

What remains to be done is to establish the relation to observables, as these do not really measure the $H(z)$ of Poisson gauge.
Observables are taken on our past light cone, which defines yet another slicing of the spacetime we live in.
In practice, a Hubble diagram is constructed by measuring
the luminosity distance $D_L(z)$ to {(more precisely the luminosity of)} far away standard candles and using the relation (for $\Omega_k=0$, for $\Omega_k\neq 0$ this relation only holds for {very small redshifts})
$$
d/dz\left[(1+z)^{-1}D_L(z)\right] = 1/H(z)\,.
$$
The perturbations of the luminosity distance have been studied at first and second order in perturbation theory~\cite{Bonvin:2005ps,Fanizza:2013doa,Umeh:2014ana,Ben-Dayan:2014swa}. These studies strongly suggest that, once the nonlinear evolution of matter has been solved non-perturbatively (e.g.\ by means of relativistic N-body
simulations like the ones presented here), the projection effects relevant for the construction of observables can be added perturbatively within the Poisson gauge
(e.g.\ by means of ray tracing). There it has been found that the presence of structure can lead to deviations in the best-fit value of $H_0$ at the 1\% level. Strong lensing will also produce a few outliers, i.e.\ individual sources that are magnified or demagnified by order-unity factors, but this effect can still be treated accurately \mkt{with $N$-body simulations and ray-tracing}, and is included in the error budget when fitting for $H_0$.

The conclusion of this work is therefore that there are time 
slicings in which the expansion rate is  relatively close to what observers measure and in these slicings backreaction is small. We used the
example of Poisson gauge, but there would be others, e.g.\ geodesic light cone gauge~\cite{Maartens:1995dx,Gasperini:2011us}. However, comoving synchronous gauge is not
well suited to describe observations in the late time clumpy universe. In the respective slicing backreaction becomes large and the gauge actually breaks down during
structure formation. Before this happens, light cone observables can of course still be worked out in this gauge, see e.g.\ \cite{Giblin:2016mjp} for a relevant
example. The projection effects are however much more intricate in this case, as several large cancellations have to occur.
\vspace{0.2cm}

\paragraph{Acknowledgments}
It is a pleasure to thank Camille Bonvin, Pierre Fleury, Ermis Mitsou as well as the organizers and participants of the workshop {\em GR effects in cosmological surveys} in Cape Town for stimulating discussions. Syksy R\"as\"anen and Timothy Clifton gave valuable comments on the preprint version of our manuscript. This work was supported by a grant from the Swiss National Supercomputing Centre (CSCS) under project IDs d45 and s710, and by the Swiss National Science Foundation.

\section*{References}
\bibliographystyle{utcaps}
\bibliography{gevolution,chris}

\end{document}